\documentclass[pre,twocolumn,showpacs,superscriptaddress,preprintnumbers,amsmath,amssymb,floatfix,10pt]{revtex4-1}

\usepackage{graphicx}  
\usepackage{dcolumn}   
\usepackage{bm}        

\newcommand{\beq}{\begin{equation}}
\newcommand{\eeq}{\end{equation}}

\newcommand{\beqa}{\begin{eqnarray}}
\newcommand{\eeqa}{\end{eqnarray}}

\newcommand \rt {\right}
\newcommand \lt {\left}
\newcommand \nline {\nonumber \\}

\newcommand{\mbfr}{{\mathbf r}}
\newcommand{\mbfG}{{\mathbf G}}

\newcommand \freem {{\cal F}}

\newcommand \pxpy[2] {\frac{\partial #1}{\partial #2}}

\newcommand \fxfy[2] {\frac{\delta #1}{\delta #2}}

\begin{document}

\title{A Phase Field Crystal Model of Phase and Microstructural Stability in Driven Nanocrystalline Systems}
\author{Nana Ofori-Opoku}
\email{oforion@mcmaster.ca}
\affiliation{Department of Materials Science and Engineering and
Brockhouse Institute for Materials Research, McMaster University,
1280 Main Street West, Hamilton, Canada L8S 4L7}

\author{Jeffrey J. Hoyt}
\affiliation{Department of Materials Science and Engineering and
Brockhouse Institute for Materials Research, McMaster University,
1280 Main Street West, Hamilton, Canada L8S 4L7}

\author{Nikolas Provatas}
\affiliation{Department of Materials Science and Engineering and
Brockhouse Institute for Materials Research, McMaster University,
1280 Main Street West, Hamilton, Canada L8S 4L7}
\affiliation{Department of Physics and Centre for the Physics of Materials, Rutherford Building, McGill University, 3600 Rue University, Montreal, Canada H3A 2T8}

\begin{abstract}
We present a phase field crystal model for driven systems which describes competing effects between thermally activated diffusional processes and those driven by externally imposed ballistic events. The model demonstrates how the mesoscopic Enrique and Bellon [Phys.~Rev.~Lett.~{\bf 84}, 2885 (2000)] model of externally induced ballistic mixing can be incorporated into the atomistic phase field crystal formalism. The combination of the two approaches results in a model capable of describing the microstructural and compositional evolution of a driven system, while incorporating elasto-plastic effects. The model is applied to the study of grain growth in nanocrystalline materials subjected to
an external driving.
\end{abstract}

\maketitle

\section{Introduction}
There has been an explosion over the past couple of decades in the investigation of nanocrystalline (NC) materials as viable replacements for components in varying industries, from microelectronics and  integrated circuits to core components in nuclear reactors. Due to the small size features exhibited by NC materials, and their propensity to have a high density of grain boundaries and phase boundaries; the mechanical, chemical, electrical and optical properties of NC materials differ quite markedly from their polycrystalline counterparts.

Whether for tuning nanostructure size features in components during their fabrication (microelectronics) or as full size components in nuclear reactors, the microstructural and phase stability of these components, when exposed to external driving, is paramount. These external processes may be attributed to, neutron or electron irradiation (in nuclear systems), ion-beam mixing (in fabrication of microelectronics) or continued plastic deformation (leading to cyclic fatigue, e.g. high energy milling)~\cite{MartinBellon97}. Of importance in these systems is the structural integrity of components. The system is driven out of equilibrium (as described by thermally activated processes, i.e., thermodynamic equilibrium), and in the thermodynamic limit, a system may exhibit chemical or structural modifications manifested in microstructural and phase alterations. The resulting long-time configurations of these systems are then dependent on the conditions responsible for external driving.

For polycrystalline materials, there are several exhaustive reviews and no shortage of literature on materials under external driving conditions, particularly when the forcing condition is a flux of energetic particles (such as electrons, ions or neutrons)~\cite{MartinBellon97,Russell84,Was89}. Several interesting and unique phenomena can occur in these materials. Examples include precipitation in under-saturated solid solutions~\cite{CauvinMartin81a,*CauvinMartin81b,*CauvinMartin82}, the lowering of the critical point in second order order-disorder transitions~\cite{Soisson92,MartinBellon97}, the patterning of dislocations~\cite{Murphy87} and the amorphization of crystalline
materials~\cite{Limoge84,MartinBellon97}. Note that external driving is also usually accompanied by the production of excess point defects, which among other things enhance thermal diffusion. The common denominator in all driven systems, is that they achieve an altered dynamical state, where the configuration of such a state is a function of the material properties and forcing conditions~\cite{MartinBellon97}. The resulting state can be rationalized as being caused by the long-time effect of two parallel yet competing mechanisms: thermally activated atomic jumps and the forced (athermal ballistic) atomic jumps arising from external forcing. Research suggests that a method to alleviate the effects of these athermal ballistic effects is the use of NC materials, which feature a very high density of potential point defect sinks such as grain boundaries or interphase boundaries. Recently, Bai {\it et al.} \cite{Bai10} have outlined a theoretical explanation for the apparent resistance of NC materials (pure Cu) to ballistic conditions. This resistance to ballistic effects has also been observed experimentally by Rose and coworkers~\cite{Rose95,*Rose97} on NC Pd and ZrO$_2$ materials, Hochbauer {\it et al.} \cite{Hochbauer05} on Cu-Nb multilayers and in various atomistic simulations \cite{Demkowicz08,Samaras02,*Samaras06}. Unfortunately, most existing atomistic models are not capable of capturing the long-time scales relevant in studying sustained ballistic effects. While techniques have been developed to accelerate atomistic simulations \cite{Bai10}, it is not clear if these methods can be extended to a time scale appropriate for the solid state diffusion mechanisms controlling the microstructural and phase stability of NC materials.

Enrique and Bellon~\cite{Enrique99,Enrique00} have recently developed a continuum description of alloys subjected to the external driving via ballistic events. The underlying theme of their model is the competing dynamics between two dynamical mechanisms: one which is thermally driven to bring the system to thermodynamic equilibrium; and the other is athermal particle exchanges driving the system out of equilibrium, an idea which dates back to Martin~\cite{Martin84}. This is accomplished through an effective free energy functional of the Cahn-Hilliard form, whereby they forgo the explicit description of discrete interactions and supersaturation of excess point defects and instead make use of effective interactions to account for the ballistic contributions. The model has been used to study compositional patterning in alloys driven by irradiation~\cite{Enrique00,Enrique01}, phase stability of alloys under irradiation~\cite{Enrique99}, non-equilibrium fluctuations in alloys under irradiation~\cite{Enrique04} and irradiation induced spinodal decomposition in the presence of dislocations~\cite{Hoyt11}. The model can capture the appropriate long-time scales needed in examining sustained ballistic effects, unfortunately it does not incorporate different crystalline orientations, grain boundaries, defects or elasto-plastic effects. These are important in the microstructural and phase stability of NC materials under sustained external forcing.

In recent years a new atomistic modelling paradigm coined the phase field crystal (PFC) method~\cite{Elder02} has emerged.  This approach is an effective modelling tool that describes microstructure evolution over diffusive time scales, while incorporating atomistic-scale elasticity, topological defects and grain boundaries. Several PFC models have been developed and used successfully in the description of solidification~\cite{Tegze09}, spinodal decomposition~\cite{Elder07}, elasto-plasticity~\cite{Stefanovic06}, thin film growth~\cite{Huang08} and structural phase transformations in pure materials~\cite{Greenwood10,*Greenwood11} and alloys~\cite{GreenwoodOfori11}. The {\it purpose of the present work} is to incorporate the mesoscale model of ballistic effects in driven systems developed by Enrique and Bellon~\cite{Enrique00} into the PFC formalism to create a simulation framework capable of describing the evolution of composition and microstructure under ballistic conditions. Like the original model, the present model does not explicitly treat the supersaturation of point defects, and instead treats the long-time ensemble average of the discrete ballistic effects as effective interactions. Our aim is to provide a modelling framework having the capabilities of simulating the long-time scales of sustained external forcing, while incorporating the atomic effects mentioned previously. We will demonstrate the properties model by examining as a first example the grain growth behaviour in the PFC formalism of NC materials under forced ballistic conditions.

The paper is organized as follows. First, we present the PFC model and its modification to account for ballistic effects. After introducing the model, we analyse the long wavelength properties of the model, where we highlight modifications to the model of Enrique and Bellon. We follow with numerical simulations of grain growth under ballistic effects. Finally we conclude and summarize.

\section{Effective PFC Energy Functional for Ballistic Mixing}
We begin with the standard binary alloy PFC energy functional in
scaled form \cite{Elder07},
\begin{align}
\freem &= \int
d\mbfr\bigg\{\frac{n}{2}\lt[B^{\ell}+B^x\lt(2\nabla^2+\nabla^4\rt)\rt]n
-\frac{t}{3}n^3\nline&+\frac{\nu}{4}n^4+\frac{\omega}{2}\psi^2+\frac{u}{4}\psi^4+\frac{K}{2}|\vec{\nabla}\psi|^2\bigg\}
\label{eldPFCalEnrgy}
\end{align}
where the energy is scaled by ${k_BT\rho_{\ell} R^d}$. The field $n$ is
the dimensionless local density, $\psi$ the dimensionless
concentration, $k_B$ is the Boltzmann constant, $T$ the temperature,
$\rho_{\ell}$ the reference liquid density, and $R$ the average atomic
radius, with $d$ being the dimensionality. Note the final three terms in Eq.~(\ref{eldPFCalEnrgy}) comprise a Cahn-Hilliard equation \cite{Cahn58,*Cahn59}, which has been used to study phase separation. $B^{\ell}$ and $B^x$ are
the dimensionless bulk moduli of the liquid and solid, respectively,
setting the energy scale of the system. Following Ref.~\cite{Elder07}, here we consider the expansion of the moduli where $B_\ell=B_0^\ell+B_2^\ell\psi^2$ and $B^x=B_0^x$. The
remaining parameters are constants, which in principle are functions
of the direct two point correlation functions, and can be calculated
from first principles or fit to phenomenological databases or
theories of materials properties, e.g. surface
energy~\cite{Wu07,Majaniemi09,Provatas10}. However, the current work will only treat these parameters as constants. In the liquid phase $n$ is
constant everywhere, while in the solid it assumes a spatially
periodic structure. In two dimensions (2D), the functional in
Eq.~(\ref{eldPFCalEnrgy}) has a phase diagram of coexisting liquid
and solid phases  with hexagonal symmetry. For greater detail in the derivation of the PFC energy function presented in Eq.~(\ref{eldPFCalEnrgy}), the reader is referred to Ref.~\cite{Elder07}.

The density and concentration fields, $n$ and $\psi$, follow dissipative dynamics that minimize $\mathcal{F}$. To include external forced ballistic effects, the dynamics are augmented by two Enrique and Bellon like source terms in the density and concentration equations. They are written as
 \begin{align}
  \pxpy{n}{t} \!&= \!M\nabla^2\fxfy{\freem}{n} \!-\!\Gamma \lt(~\!n\! - \langle\,n\,\rangle_{\cal{R}}\,\rt)
\label{eldPfcDyn_dens}
\end{align}
\begin{align}
\pxpy{\psi}{t} \!&=\! M\nabla^2\fxfy{\freem}{\psi}-\!\Gamma\lt(~\! \psi \!-\! \langle\,\psi\,\rangle_{\cal{R}}\, \rt)
\label{eldPfcDyn_conc}
\end{align}
with $M$ the mobility and $\Gamma$ the frequency of forced atomic exchanges, proportional to the flux of incident particles. Note that Eqs.~(\ref{eldPfcDyn_dens}) and (\ref{eldPfcDyn_conc}) are deterministic, however to incorporate the short time fluctuations of the system, in Section~\ref{dynamics} when we report on numerical simulations, we append to these equations two stochastic variables, $\zeta$ and $\xi$, having a Gaussian distribution with zero mean and amplitudes $A_{\zeta}$ and $A_{\xi}$  for the density and concentration fields, respectively. $\langle\,\psi\,\rangle_{\cal{R}}$ and $\langle\,n\,\rangle_{\cal{R}}$ denote the corresponding weighted spatial averages of the concentration and density fields respectively, in response to forced atomic exchanges with an average distribution distance ${\cal{R}}$. The weighted averages are defined by,
\begin{align}
\langle\,\psi\,\rangle_{\cal{R}} = \int d\mbfr^\prime w_{\cal{R}}(\mbfr \!-\! \mbfr^\prime)\psi(\mbfr^\prime)
\end{align}
and
\begin{align}
\langle\,n\,\rangle_{\cal{R}} = \int d\mbfr^\prime w_{\cal{R}}(\mbfr \!-\! \mbfr^\prime)n(\mbfr^\prime).
\end{align}
The function $w_{\cal{R}}$ is a weight function describing the long-time spatial extent of ballistic exchanges in crystalline materials. A Yukawa potential was chosen as a form for the weight function in Ref.\cite{Enrique00}, for the case of exchanges driven by irradiation. However in the present work we adopt the 2D analog;
\begin{align}
w_{\cal{R}}(r) = \frac{1}{2\pi {\cal{R}}^2}K_0\lt(\frac{r}{{\cal{R}}}\rt),
\label{weightFunc}
\end{align}
where $K_0$ is a modified Bessel function of the second kind. As stated above, ${\cal{R}}$ describes the average spatial extent of ballistic events and is also related to the energy of the ballistic driving force. We note that the above is the lowest order description of ballistic events in driven systems that can be included for a binary alloy model that has coupling between density and composition variations. In general $\Gamma$ can take on different values for the density and concentration field, respectively, and two separate functions $w_{\cal{R}}^{n}$ and $w_{\cal{R}}^{\psi}$ can be introduced for each respective field. The model presented here, like the model of Enrique and Bellon, has the underlying idea that a system exposed to external driving can be described by two mechanisms acting in parallel: thermal diffusion described by terms of the form $\!M\nabla^2\fxfy{\freem}{n}$ and ballistic events described by terms of the form $\Gamma \lt(~\!n\! - \langle\,n\,\rangle_{\cal{R}}\rt)$. Similarly for the concentration order parameter. Note that when ${\cal{R}}=0$, dynamics are driven only by the thermal portions of Eqs. (\ref{eldPfcDyn_dens}) and (\ref{eldPfcDyn_conc}).

Including only the lowest level of description, we construct an effective PFC energy functional for ballistic events, which affords an opportunity to explore equilibrium properties under ballistic conditions. Following Enrique and Bellon, we can write the effective functional as $E=\freem +\gamma \mathfrak{G}$, where $\mathfrak{G}$ is the external energy source representing effective interactions arising due to ballistic effects and $\gamma=\Gamma/M$, the reduced forcing frequency. The effective interactions are described by a self interaction energy written as
\begin{align}
\mathfrak{G} \!=\!
\frac{1}{2}\int \! d\mbfr \, d \mbfr^\prime~\big\{
n(\mbfr)g(\mbfr &\!-\! \mbfr^\prime)n(\mbfr^\prime)+\psi(\mbfr)g(\mbfr \!-\! \mbfr^\prime)\psi(\mbfr^\prime)\big\},
\label{slfIntAlloy}
\end{align}
where the kernel for ballistic exchanges $g$ satisfies the equation
$\nabla^2g(\mbfr\!-\!\mbfr^\prime) \!=\! -\lt[\delta(\mbfr\!-\!\mbfr^\prime) \!-\! w_{\cal{R}}(\mbfr\!-\!\mbfr^\prime)\rt]$. Which is to say, the kernel satisfies the Poisson solution to a point source perturbed by a weight function.

\section{Long Wavelength and Equilibrium Properties}
We first interrogate the long wavelength (i.e., phase field limit) limit of the model.
Secondly, we examine its bulk properties under forced driving, i.e., a kinetic phase diagram under ballistic conditions. To examine the long wavelength properties of the model, we perform an amplitude expansion of our effective PFC free energy, $E$. To perform the expansion, we will be substituting into the effective free energy a single mode expansion for the density,
\begin{equation}
n=n_o(\mbfr)+\sum_{\{\mbfG_{j},j\ne 0\}}\eta_j(\mbfr)\,e^{i\mbfG_{j}\cdot\mbfr}+\text{c.c}.
\end{equation}
 $n_o(\mbfr)$ is the dimensionless average density of the system, a conserved quantity which is slowly varying spatially with respect to $\mbfG_{j}$, $\eta_j(\mbfr)$ are dimensionless spatially varying complex amplitudes which like the average density are slowing varying on atomic scales, $\mbfG_{j}$ represents the lowest set of reciprocal basis vectors necessary to describe the crystal structure of interest and \text{c.c.} denotes the complex conjugate. The result is then {\it coarse grained} using the volume averaging method of Refs.~\cite{Majaniemi09} and \cite{Provatas10}. For a crystal of hexagonal symmetry in 2D, coarse graining $E$ gives, to lowest order in the ballistic terms;
\begin{align} \tilde{E} &= \int~d\mbfr~ \Bigg\{ \sum_{j=1}^3\lt(\Delta B_0+
\gamma\hat{g}(|\mbfG_{j}|,{\cal{R}}) +3\nu n_o^2-2tn_o\rt)\,|\eta_j|^2
\nline
&+\sum_{j=1}^3\lt[B_0^x\,|{\cal G}_j\,\eta_j|^2 +
\frac{3\nu}{2}|\eta_j|^4\rt]+(6\nu n_o-2t)\lt(\prod_{j=1}^3\eta_j +
\text{c.c.}\rt) \nline
&+6\nu\sum_{j,k> j}^{3}|\eta_j|^2|\eta_k|^2
+\lt(\omega+2B_2^\ell\sum_{j=1}^3|\eta_j|^2\rt)\frac{\psi^2}{2}
+\frac{u}{4}\psi^4\nline
&+\frac{K}{2}|\vec{\nabla}\psi|^2
+\lt(\frac{\Delta
B_0}{2}+\frac{B_2^\ell\,\psi^2}{4}\rt)n_o^2-\frac{t}{3}\,n_o^3+\frac{\nu}{4}\,n_o^4\nline&+\frac{B_0^x}{2}\lt(\lt[1+\nabla^2\rt]n_o\rt)^2
+\gamma\frac{n_o}{2}\int
d\mbfr^\prime~g(\mbfr-\mbfr^\prime)n_o(\mbfr^\prime)\nline
&+\gamma\frac{\psi}{2}\int d\mbfr^\prime
g(\mbfr-\mbfr^\prime)\psi(\mbfr^\prime)\Bigg\},
\label{ampExpAlloy}
\end{align}
where $\Delta B_0=B_0^\ell-B_0^x$ defines a temperature scale, the caret denotes the Fourier transform, with $\hat{g}={\cal{R}}^2/(1+|\mbfG_{j}|^2{\cal{R}}^2)$ and ${\cal G}_j\equiv \nabla^2 + 2i\mbfG_{j}\cdot\mathbf{\nabla}$. The free energy functional of Eq.~(\ref{ampExpAlloy}) describes a system with multiple crystal orientations, elasto-plastic effects through the complex nature of the amplitudes, as well as density changes through the field $n_o$. Moreover, as a coarse grained model, it operates on scales much larger than the lattice constant of the solid. A noteworthy result of Eq.~(\ref{ampExpAlloy}) is that the $\gamma\hat{g}(|\mbfG_{j}|,\cal{R})$ term competes with the temperature ($\Delta B_0$), as both are modulated by the amplitude magnitude squared, $|\eta_j|^2$. In other words forced external driving is manifested, partly, as shifts in the temperature scale leading to a new effective temperature. This is in agreement with previous theories of phase equilibria under driven conditions~\cite{Martin84,Soisson92}. Furthermore, the ballistic term in concentration, $\gamma\frac{\psi}{2}\int d\mbfr^\prime g(\mbfr-\mbfr^\prime)\psi(\mbfr^\prime)$, will be shown to renormalize the coefficient of the $\psi^2$ in the Cahn-Hilliard portion of the energy in the thermodynamic limit, i.e., limit ${\cal R}\rightarrow\infty$. Namely, the model also predicts that ballistic effects cause alterations in the critical transition temperature as well. We also note that in the absence of density variations, defects and the elasto-plastic effects inherent in the complex nature of the amplitude formulation, i.e., all $|\eta_j|=\text{const.}$ and scale out of the problem, we recover the Cahn-Hilliard ballistic model of Enrique and Bellon~\cite{Enrique00}. The repercussions of these effects are explored below when we construct phase diagrams for this model.

\subsection{Pure Material}
\begin{figure}[htbp]
\resizebox{3.5in}{!}{\includegraphics{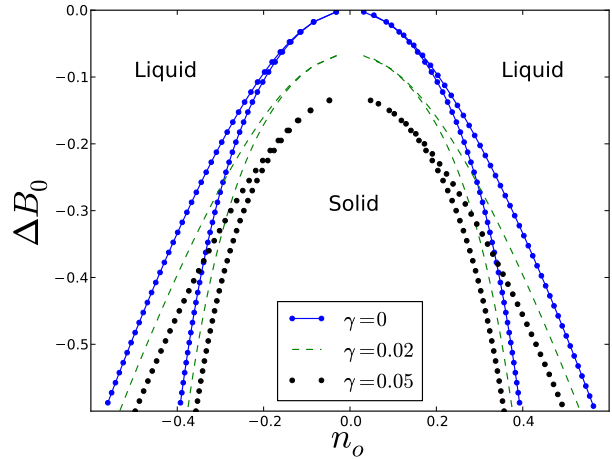}}
\caption{(colour online) Kinetic phase diagram for a pure material exhibiting density variations in $\Delta B_0\times n_o\times\gamma$ space, shown here projected onto the $\Delta B_0\times n_o$ space for choice values of the reduced forcing frequency $\gamma$. Squares with lines(blue) represent the standard PFC model, $\gamma=0$, dashed lines(green) correspond to a ballistic case of $\gamma=0.02$, while squares(black) correspond to a ballistic case of $\gamma=0.05$.}
\label{fig:iradPFC_pure}
\end{figure}

The equilibrium phase diagrams follow standard minimization procedures analogous to Refs.~\cite{Elder02} and \cite{Elder07}. For a pure material, we assume a real and constant amplitude  $\eta_j\equiv\phi$ for all $j$ and a constant wave vector, to describe the periodicity of the solid, ignore all terms in $\psi$ and set $t=0$ and $\nu=1$. Making a long wavelength approximation, i.e., $q{\cal{R} } \gg 1$, we first begin by minimizing the energy for the equilibrium wave vector, $q_{\text{eq}}=|\mbfG_j|$, which is substituted back into the energy. Next, the energy is minimized for the real amplitude $\phi$. After substitution of the minimized amplitude back into the energy, we have a resulting energy which is a function of the conserved quantity $n_o$, temperature ($\Delta B_0$), and the reduced forcing frequency $\gamma$. A common tangent or the equivalent Maxwell equal area construction leads to Fig.~\ref{fig:iradPFC_pure}, which plots the a dynamical phase diagram in $\Delta B_0$, $n_o$ and $\gamma$ space. We notice a trend of decreasing solid-liquid transition temperature with increasing $\gamma$. As previously discussed, we can attribute this behaviour to the competition between the temperature and ballistic term, where in the thermodynamic, constant amplitude limit, the competition term takes the form of $(\Delta B_0 + \gamma/q)\phi^2$.

\subsection{Eutectic Binary Alloy}
\begin{figure}[htbp]
\resizebox{3.5in}{!}{\includegraphics{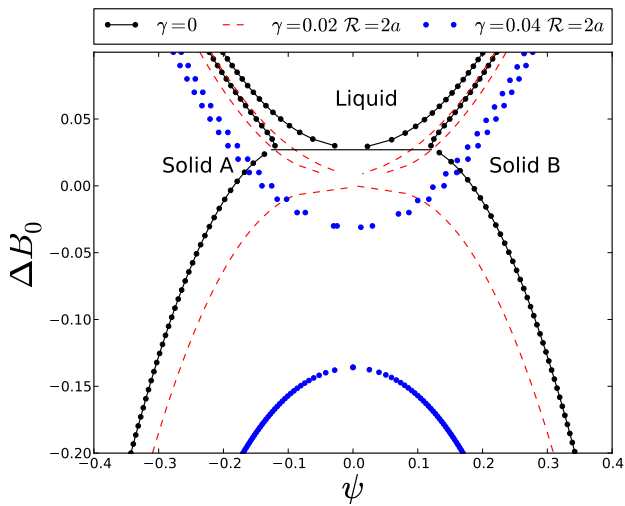}}
\caption{(colour online) Binary eutectic phase diagram under forced ballistic driving. circles and lines(black) represent the standard binary PFC model, $\gamma=0$, dashed lines (red) correspond to a ballistic case of $\gamma=0.02$ and ${\cal{R}}=2a$, while circles(blue) correspond to a ballistic case of $\gamma=0.04$ and ${\cal{R}}=2a$ with $a$ being the lattice spacing.}
\label{fig:iradPFC_alloy}
\end{figure}

Moving on to the binary alloy, we again assume real and constant amplitude  $\eta_j\equiv\phi$ for all $j$ and a constant wave vector, $q$, to describe the periodicity of the solid. We set $B_2^\ell=-1.8$, $B_0^x=1$, $t=0.6$, $\nu=1$, $u=4$ and $n_o=0$. A spinodal and a eutectic alloy are differentiated, in this PFC model, by choice of the parameter $\omega$. Here we report the calculation for the eutectic alloy only, i.e. $\omega=0.008$. After the parameters are set, we follow the same minimization steps as outlined for the pure material in the preceding section with respect to $\phi$ and $q_{\text{eq}}$. With the average density being chosen to be zero, the resulting minimized energy is a only a function of $\psi$, temperature ($\Delta B_0$), reduced forcing frequency $\gamma$ and ballistic distance ${\cal{R}}$. A common tangent construction or equivalent Maxwell equal area construction gives Fig.~\ref{fig:iradPFC_alloy}. In Fig.~\ref{fig:iradPFC_alloy}, we present the phase diagram of a eutectic alloy under ballistic effects in $\Delta B_0$, $\psi$ and $\gamma$ phase space. In the figure, $a=2\pi/q_{\text{eq}}$ is the lattice spacing associated with the hexagonal unit cell in 2D. There are several things to note. There is still a general shift of the solid-liquid transition temperature, additionally however, we also witness the alteration of the eutectic alloy into a spinodal alloy. Again, in the thermodynamic limit the coefficient of the amplitude square term becomes $(\Delta B_0 + \gamma/q)\phi^2$, which from the pure material calculation above, was shown to affect the solid-liquid transition. This term is what primarily leads to the general shift in the solid-liquid transition. However, for the binary alloy, the coefficient multiplying the $\psi^2$ in the Cahn-Hilliard contribution becomes $\lt(\omega+12B_2^\ell\phi^2 + \gamma {\cal{R}}^2\rt){\psi^2}$. The $\gamma {\cal{R}}^2$ (i.e. $q{\cal{R} } \gg 1$) contribution to this term simultaneously accounts for the change from eutectic to spinodal, the ever increasing range of the solid solution region and contributes to the change in the solid-liquid transition lines in our model as well.

\section{Dynamics}
\label{dynamics}
Two sets of numerical simulations were conducted to study the role of ballistic driving on the grain growth of a pure nanocrystalline material. The first is designed to illustrate the physics of the proposed ballistic PFC term. The second set of simulations examines the grain growth behaviour of a polycrystalline sample under ballistic conditions. For the simulations, we solve Eq.~(\ref{eldPfcDyn_dens}), with $\Delta{x}=0.785$, $\Delta{t}=1$, $M=1$, $\Delta{B_0}=-0.26$, $n_o=0.285$ and $\zeta\ne0$ (with amplitude $A_{\zeta}=0.01$), using a semi-implicit Fourier technique.

\subsection{Coarsening of a 3-Sided Grain}
\begin{figure}[htbp]
    \centering
    \includegraphics[width=3.5in]{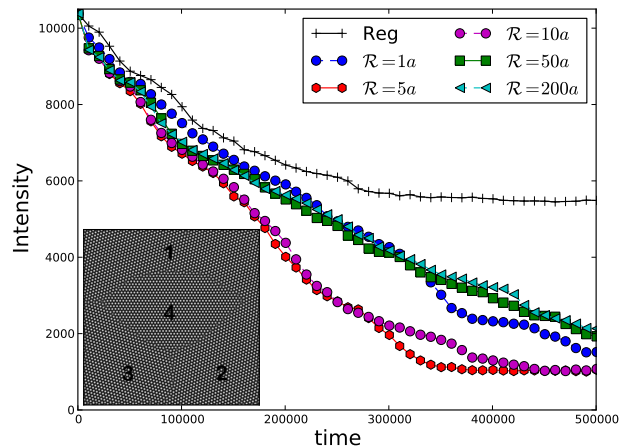}
     \caption[]{(colour online) Grain intensity versus time for a $512\Delta x \times512\Delta x$, with $\gamma=0.01$ for varying varying ${\cal{R}}$. Left bottom inset shows an example of the initial microstructure used to perform the simulations. Density ($n$) map is shown, where lighter regions represent areas with a higher probability.}
 \label{fig:dyn3sidegrain}
\end{figure}

For the first set of simulations, we have $4$ grains as shown in the inset of Fig.~\ref{fig:dyn3sidegrain} as an initial structure. In this configuration, conventional theory states that grains ``$1$'', ``$2$'' and ``$3$'' will tend to grow at the expense of grain ``$4$'', driven by gradients in the chemical potential resulting from curvature effects. Simulations were performed for the regular PFC dynamics ($\gamma=0$), for the proposed ballistic PFC model at $\gamma=0.01$ and for several values of the ballistic distance, i.e. ${\cal{R}}=1a, 5a, 10a, 50a$ and $200a$. For a particular measure of the coarsening rate, we focus on the behaviour of grain ``$4$''. Results were averaged over several runs, where grains ``$1$'', ``$2$'' and ``$3$'' assumed different orientations while keeping grain ``$4$'' fixed. In Fig.~\ref{fig:dyn3sidegrain}, we plot the average intensity of the Bragg peaks obtained from the power spectrum for grain ``$4$'' (which is proportional to its area) versus time. Several interesting aspects of the plot are noteworthy. First, the rate of coarsening is enhanced for all cases of ballistic mixing. However there exists two regimes of noteworthy behaviour. Namely, we find one regime wherein the ${\cal{R}}= 5a$ and $10a$ conditions cluster and another for the ${\cal{R}}=1a, 50a$ and $200a$ conditions. In the first regime we have the fastest coarsening rates.

The results of Fig.~\ref{fig:dyn3sidegrain} can be understood as follows. The long-time ballistic effect acts over a length scale $\mathcal{R}$ and within this spatial extent the mobility and/or driving force is enhanced. Thus, when $\mathcal{R}$ becomes comparable to the size of grain ``4'', the coarsening rate is more pronounced as seen in the ${\cal{R}}=5a$ and $10a$ cases. The plateaus evident at late time for these two conditions are a result of the background signal of the spectrum after the disappearance of grain ``$4$'' from the system. When the ballistic distance is much larger than the size of grain ``4'', i.e., ${\cal{R}}=50a$ and $200a$ in the second regime, there is equal enhancement in the competition from all grains for growth (although still enhanced compared to the $\gamma=0$ case). This results in a deadlock, which does not allow for preferential growth. Concerning the ${\cal{R}}=1a$, the spatial extent of this ballistic event is not large enough to impact any relevant length scales necessary for pronounced preferential growth. However having influence on all interatomic length scales in the system, particularly those associated with boundary widths, it effectively behaves like the larger ballistic distances.

\subsection{Grain Growth}
\begin{figure}[htbp]
    \centering
    \begin{tabular}{c}
    \includegraphics[width=3.5in]{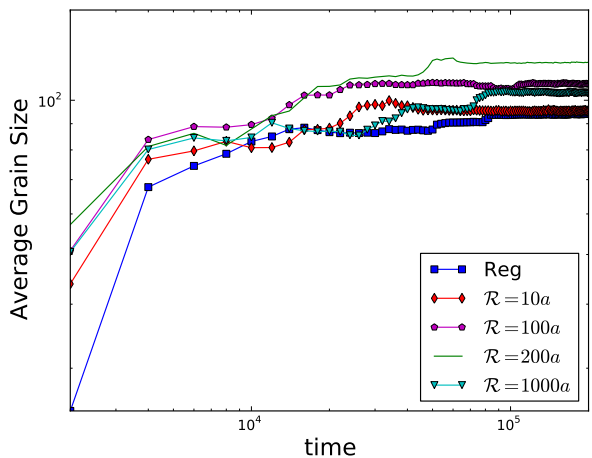}
    \\
    \end{tabular}
    \caption[]{(colour online) Grain growth results plotting average grain size versus time, on a log-log scale, for a $1024\Delta x \times1024\Delta x$. Here $\gamma=0.05$ for ${\cal{R}}=10a, 100a, 200a$ and ${\cal{R}}=1000a$.}
 \label{fig:graingrowth}
\end{figure}

The second set of simulations examines grain growth for a nanocrystalline sample in a $1024\Delta x \times1024\Delta x$ system, with approximately $100$ grains. Figure~\ref{fig:graingrowth} plots the average grain size versus time, for $\gamma=0.05$ for varying ballistic distances ${\cal{R}}$. Grain growth was quantified by Fourier analysis, where the full width at half maximum of the the Bragg peak was measured in the radially averaged signal. The multigrain simulations are consistent with the results of Fig.~\ref{fig:dyn3sidegrain}. Grain growth is enhanced for all ${\cal{R}}$ values. Also note, in the time range of $500-1000$, the average grain size is $~75a$, and the growth is enhanced most strongly for ${\cal{R}}$ comparable to this grain size (${\cal{R}}=100a$). Therefore, our results demonstrate that nanocrystalline materials, although possibly offering increased resistance when exposed to ballistic driving forces, such as that from irradiation damage, are susceptible to enhanced grain growth under such driving forces, with the rate of growth generally increasing with increasing ballistic energy (i.e.  ${\cal{R}}$), and particularly enhanced when {$\cal{R}$} is comparable to the grain size.

\section{Summary and Conclusions}
In summary, we have developed an atomistic continuum model of competing effects between thermally activated diffusional processes and those driven by externally imposed ballistic events in both pure materials and alloys using a PFC free energy functional. As an extension of the model of Enrique and Bellon, our approach is capable of describing defect microstructure, multiple crystal orientations, and elasto-plastic effects. A long wavelength analysis of our model predicts that ballistic events change the phase diagram through a shift in the solid-liquid transition temperatures for a pure material. For a eutectic alloy, the model further predicts an alteration of equilibrium through shifts in the critical transition temperatures in the solid solution regions, an effect which have long been conjectured in the irradiation damage literature. Through an investigation of the kinetic properties of the model, we find that under ballistic driving, grain growth is enhanced with increasing average ballistic distance, with a predicted enhancement of grain sizes comparable to the ballistic range $\cal{R}$. Further numerical and experimental investigation is required to elucidate the possible myriad of behavioural aspects in nanocrystalline system when subjected to ballistic driving.

We acknowledge the National Science and Engineering Research Council of Canada
(NSERC) for financial support and David Montiel and Jonathan Stolle for useful discussions.

%


\end{document}